\documentclass[aps,pre,amsmath,amssymb,twocolumn,notitlepage,longbibliography,superscriptaddress]{revtex4-2}
\usepackage{graphicx} % Required for inserting images
\usepackage[colorlinks=true,allcolors=blue]{hyperref}
\usepackage{xcolor}
\usepackage{amsmath}
\usepackage{amsfonts}
\usepackage{amssymb}
\usepackage{varioref}
\usepackage{float}
\usepackage{dcolumn}   % needed for some tables
\usepackage{subfigure}
\usepackage{bm}% bold math
\usepackage{multirow}
\usepackage[normalem]{ulem}
\usepackage[type={CC},modifier={by},version={4.0}]{doclicense}

\begin{document}
\title[]{Non-linear visco-elasto-plastic rheology of a viscous vertex model}
\thanks{\doclicenseThis}

\author{Shalabh K. Anand}
\email{shalabh.anand@uibk.ac.at}
\affiliation{Aix Marseille Univ, Université de Toulon, CNRS, CPT (UMR 7332), Turing Centre for Living systems, Marseille, France.}
\affiliation{Institut für Theoretische Physik, Universität Innsbruck, Technikerstrasse 21A, A-6020, Innsbruck, Austria}
\author{Matthias Merkel}
\email{matthias.merkel@univ-amu.fr}
\affiliation{Aix Marseille Univ, Université de Toulon, CNRS, CPT (UMR 7332), Turing Centre for Living systems, Marseille, France.}

\begin{abstract}
	Morphogenesis involves complex shape changes of biological tissues.
	Yet, tissue shape changes depend on tissue rheology, which in turn arises from the interplay of large numbers of cells. 
	Here, we link cell- and tissue-scale mechanics by constructing mean-field rheological relations for the vertex model.
	In contrast to past work in the field, we study a vertex model with an explicit viscous friction.
	We also include two different cellular mechanisms creating active, anisotropic stresses.
	Our mean-field model accounts for cell shape and the non-linear elastic and visco-plastic regimes.
	We validate our results by predicting the response to large-amplitude oscillatory shear.
	There are several vertex model variants, and comparing to results from the literature, we show that their rheology depends on a number of model details. 
	Our approach should be sufficiently general to construct non-linear mean-field constitutive relations for any cell-based tissue model.
\end{abstract}

\maketitle

\section{Introduction}
Understanding the mechanical behavior of biological tissues is crucial for deciphering complex processes in developmental biology, tissue engineering, and cancer metastasis. 
Yet, how tissue-scale mechanics arises from the collective behavior of many cells is subject to active research, including experimental work \cite{Harris2012,Serwane2017,Mongera2018,Khalilgharibi2019,Gomez-Gonzalez2020,Tlili2022,Guillamat2025}, theoretical work \cite{Bi2015,Popovic2017a,Merkel2019,Bonfanti2020}, and both combined \cite{Etournay2015,Petridou2021,Serra2023,Dessalles2025}.

To theoretically study how tissue dynamics depends on cellular mechanics, one needs a tissue model where cells are explicitly represented \cite{Merkel2017b}. Such models include particle-based models \cite{Matoz-Fernandez2016,Germann2019a}, the Cellular Potts Model \cite{Graner1992,Belousov2024}, the vertex model \cite{Honda1984,Farhadifar2007,Alt2017}, and phase-field models \cite{Mueller2019,Loewe2020}. In order to achieve a good compromise between the ability to represent features such as cell shape, not too unrealistic time-dependent dynamics, and reasonable computation time, we use here the vertex model.

Previous work has already studied some aspects of vertex model rheology, including its linear rheology \cite{Duclut2022,Damavandi2025,Tong2022a,Tong2023}.
Among those, Tong et al.\ \cite{Tong2023} analytically connect tissue-scale linear rheology to vertex model details using the elegant normal mode formalism, which has also been applied to particle-based tissue models \cite{Henkes2020}.

Further work has focused on tissue-scale elastic properties. For instance, prominently, several vertex model variants are known to exhibit a rigidity transition, which was studied for both ordered \cite{Farhadifar2007,Staple2010} and disordered \cite{Bi2015} packings.
This transition, and the related linear and non-linear elastic properties, have subsequently been studied in a series of papers \cite{Moshe2017,Sussman2018,Merkel2018,Merkel2019,Wang2020,Huang2022,Hernandez2022,Lee2024,Lee2024a,Kim2024,Kim2024b}. However, stress dissipation and the non-linear visco-plastic regime of the vertex model is less well explored.

In the vertex model, there are several ways to implement mechanical stress dissipation on the cell scale.
Most often, a substrate friction is implemented, where the frictional force on a vertex is proportional to the vertex velocity.
In this case, to probe rheology, care needs to be taken to use the vertex velocity \emph{relative} to a deforming substrate \cite{Tong2023}.
However, many tissues do not have any substrate -- take for instance 3D stem cell aggregates \cite{Gsell2025}, where friction is dominated by internal dissipative processes such as viscous friction.
To properly describe such cases in a vertex model, a friction is needed that is Galilean invariant: whether a tissue is moving at some constant velocity or not, nothing should change about the forces the cells experience.
Therefore, several models include friction terms that only depend on velocity vector \emph{differences} \cite{Okuda2015,Tlili2019,Tong2023,Rozman2025,Koyama2026}.
Yet, without a substrate and in the over-damped limit, one can also globally \emph{rotate} a tissue at constant rate without any relevant additional force.
In other words, appropriate models should also display a ``Galilean invariance'' with respect to \emph{rotation}. 
Both invariances should be required to describe any tissue-internal dissipation processes.
Yet, as far as we know, very little work so far satisfies both invariances for their internal frictions \cite{Brodland2007,Staple2012,Nestor-Bergmann2018,Fu2024,Lin2026}.
Note that, as an alternative to explicitly taking dissipative processes into account, also quasi-static relaxation obeys these two invariances \cite{Merkel2014b,Duclut2021,Duclut2022}.
In this article, we will explicitly account for the internal dissipation generated in cell-cell interfaces, cell perimeters, and cell areas, broadly following ideas from Ref.~\cite{Staple2012}.

On the tissue scale and at long times, mechanical stresses relax mostly through T1 transitions. This was studied in the vertex model, where T1 transitions can be triggered by externally applied stresses \cite{Bi2014,Popovic2021,Kim2021a,Guigue2025}, active fluctuations \cite{Bi2016,Duclut2021}, or thermal noise \cite{Sussman2018b,Li2021a,Li2025,Pandey2025,Li2026}.
Here, to systematically characterize the tissue-scale stress relaxation due to T1 transitions, we will employ a triangle-based method to decompose shear of a cellular network into contributions by cell shape change and T1 transitions \cite{Merkel2017}. This approach was previously used by one of us to characterize tissue-scale rheology \cite{Merkel2014b}, but for a vertex model without explicit internal dissipation.
One of us also applied a related approach to emulsion experiments, where we observed quasi-static, plastic stress relaxation.
In these emulsions, much of the plastic behavior could be understood in terms of a reversible fraction function \cite{Guigue2025}.
Here, we will follow a similar approach, but for a vertex model with explicit internal dissipation.

Many vertex models also include active terms \cite{Barton2017,Krajnc2018,Sknepnek2023,Lin2023,Rozman2024a,Damavandi2025}.
Notably, we recently studied two different cellular mechanisms of creating active, anisotropic stresses in biological tissues, motivated by active deformations observed in frog embryo explants \cite{Barrett2025}. Intriguingly, we found that cell shape $Q$ behaved differently in both cases, calling for a mean-field model of non-linear vertex model rheology that explicitly accounts for cell shape.

Here, we implement a vertex model with explicit frictional dynamics that fulfills translational and rotational Galilean invariance.
We show how, using the strain decomposition from Ref.~\cite{Merkel2017}, tissue shear rheology, i.e.\ the complex relation between shear stress $\tilde\sigma$ and shear rate $\tilde{v}$ can be separated into two relations: (i) shear stress depending on cell shape and shear rate, $\tilde\sigma(Q, \tilde{v})$, and (ii) the shear rate contribution by T1 transitions, $R(Q, \tilde{v})$.
To quantify these relations, we first use small-amplitude oscillatory pure shear simulations to study the linear rheology.
Subsequently, we use constant pure shear rate simulations, to include effects from non-linear elasticity, visco-plastic stress relaxation by T1 transitions, and activity.
Our approach can be used to develop a mean-field rheology for any vertex model variant.

\section{Methods}
\subsection{Vertex model}
The 2D vertex model describes an epithelial tissue as a tiling of polygons, each of which represents a cell.
The state of the model depends on the positions $\bm{r}_i$ of the polygon corners, called vertices, which will be denoted by Latin indices $i,j,\dots$ throughout this article.
Elastic forces on the vertices,  $\bm{F}^\mathrm{el}_i=-\partial E/\partial\bm{r}_i$, are defined by an energy functional $E(\lbrace\bm{r}_i\rbrace)$. 
We use a classical form for the energy functional \cite{Farhadifar2007,Staple2010}, which reads in dimensionless form:
\begin{equation}
  E = \frac{1}{2} \sum_{c=1}^N{\bigg[k_A(a_c-1)^2 + (p_c-p_0)^2\bigg]}.
  \label{eq:E}
\end{equation}
Here, the sum is over over all $N$ cells $c$.
The first term in the sum represents area elasticity of cell $c$, where $k_A$ is an elastic modulus and $a_c$ the cell area. 
The second term represents cell perimeter elasticity, where $p_c$ denotes the perimeter of cell $c$, and $p_0$ is a preferred perimeter.

We use periodic boundary conditions with periodic box dimensions $L_x\times L_y$.
We impose time-dependent pure shear, $\gamma(t)$, such that $L_x(t) = L_{x0}e^{\gamma(t)}$ and $L_y(t) = L_{y0}e^{-\gamma(t)}$ with $L_{x0}L_{y0}=N$ in dimensionless units. Note that this dynamics keeps the total box area constant with $L_xL_y=N$.

For the time-dependent dynamics of the vertex model, we focus on the over-damped limit, which implies force balance on each vertex $i$:
\begin{equation}
	0 = \bm{F}^\mathrm{el}_i + \bm{F}^\mathrm{fr}_i + \bm{F}^\mathrm{act}_i.
	\label{eq:force balance}
\end{equation}
Here, $\bm{F}^\mathrm{el}_i$ is the elastic force defined right above Eq.~\eqref{eq:E}, $\bm{F}^\mathrm{fr}_i$ are friction forces, and  $\bm{F}^\mathrm{act}_i$ are active forces.
Since many tissues or parts of tissues are not directly in contact with any solid substrate, we want to study a model that fulfills translational and rotational Galilean invariances. 
This corresponds to the local conservation of linear and angular momentum.
We note that the elastic forces $\bm{F}^\mathrm{el}_i$ already fulfill these invariances. 

To introduce friction forces $\bm{F}^\mathrm{fr}_i$ that fulfill translational and rotational Galilean invariances, we introduce scalar frictional elements.
In this article, we introduce frictions on all (i) cell areas, (ii) cell perimeters, and (iii) cell-cell interface lengths.
Specifically, we introduce 
(i) an area stress $\sigma^a_c=(\bar\eta^a/a_c)\dot{a}_c$ for any cell $c$, where the dot indicates the time derivative,
(ii) a perimeter tension $T^p_c=(\bar\eta^p/p_c)\dot{p}_c$ for any cell $c$, and
(iii) an interface tension $T^\ell_{ij}=\bar\eta^\ell\dot{\ell}_{ij}$ for any interface of length $\ell_{ij}$ between any two connected vertices $i$ and $j$.
Following Ref.~\cite{Staple2012}, we obtain for the friction forces:
\begin{equation}
	\bm{F}^\mathrm{fr}_i = -\sum_{m=1}^{N_\mathrm{fr}}{\frac{\partial g_m}{\partial\bm{r}_i}\eta_m \dot{g}_m},
	\label{eq:Ffr}
\end{equation}
where $m$ goes over all $N_\mathrm{fr}=5N$ frictional elements $g_m$.
These include:
(i) all $N$ cell areas $a_c$, where, for $m=c=1,\dots,N$, we set $g_m=a_c$, $\eta_m=\bar\eta^a/a_c$, and thus $\eta_m\dot{g}_m=\sigma^a_c$; 
(ii) all $N$ cell perimeters $p_c$, where, for $m=N+c=N+1,\dots,2N$, we set $g_m=p_c$, $\eta_m=\bar\eta^p/p_c$, and thus $\eta_m\dot{g}_m=T^p_c$; 
(iii) all $3N$ cell-cell interface lengths $\ell_{ij}$, where, for $m=2N+1,\dots,5N$, we set $g_m=\ell_{ij}$, $\eta_m=\bar\eta^\ell$, and thus $\eta_m\dot{g}_m=T^\ell_{ij}$.
Furthermore, we use that all $\dot{g}_m$ can be expressed in terms of vertex movements and changes of the boundary dimensions, i.e.\ of the pure shear variable $\gamma$:  $\dot{g}_m=(\partial g_m/\partial\gamma)\dot{\gamma} + \sum_j{(\partial g_m/\partial\bm{r}_j)\cdot\dot{\bm{r}}_j}$.
We thus have:
\begin{equation}
	F^\mathrm{fr}_{i\alpha} = -\sum_{m,j,\beta}{\frac{\partial g_m}{\partial r_{i\alpha}}\eta_m \frac{\partial g_m}{\partial r_{j\beta}}}\dot{r}_{j\beta} - \sum_{m}\frac{\partial g_m}{\partial r_{i\alpha}}\eta_{m}\frac{\partial g_m}{\partial\gamma}\dot{\gamma},
	\label{eq:Ffr-2}
\end{equation}
where here and throughout this article, Greek letters denote dimension indices, $\alpha,\beta,\dots\in\lbrace x, y\rbrace$. The first term in  Eq.~\eqref{eq:Ffr-2} corresponds to what has been obtained previously \cite{Staple2012,Nestor-Bergmann2018}. The additional second term that we obtain here is due to the shear of the periodic box.

\begin{figure}[t]
	\centering
	\includegraphics[width=\linewidth]{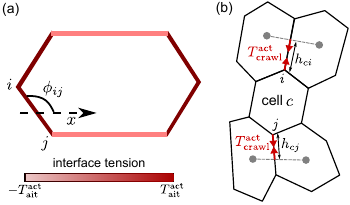}
	\caption{Schematic showing the implementation of the two active cellular mechanisms:
	(a) anisotropic interface tension and 
	(b) active crawling forces.
	}
	\label{fig:model}
\end{figure}

To ensure that also the active forces $\bm{F}^\mathrm{act}_i$ fulfill translational and rotational Galilean invariance, we use a similar form as for the friction force, Eq.~\eqref{eq:Ffr}:
\begin{equation}
	\bm{F}^\mathrm{act}_i = -\sum_{n=1}^{N_\mathrm{act}}{\frac{\partial h_n}{\partial\bm{r}_i}T^\mathrm{act}_n},
	\label{eq:Fact}
\end{equation}
where the sum is over all $N_\mathrm{act}$ active elements $h_n$, each of which experiences an active tension $T^\mathrm{act}_n$.
In this article, we include two different types of activity \cite{Barrett2025} (\autoref{fig:model}):
(a) We study anisotropic interface tensions (\autoref{fig:model}a), where the active elements are the $N_\mathrm{act}=3N$ cell-cell interface lengths $h_{ij}=\ell_{ij}$ with $n\equiv ij$, between vertices $i$ and $j$, respectively. For each interface $ij$, we introduce an active tension $T^\mathrm{act}_{ij}$ that depends on the absolute angle $\phi_{ij}$ of the interface with respect to the $x$ axis:
\begin{equation}
	T^\mathrm{act}_{ij} = -T_\mathrm{ait}^\mathrm{act} \cos{2\phi_{ij}}
	\label{eq:tensions ait}
\end{equation}
Here, $T_\mathrm{ait}^\mathrm{act}$ is a parameter that scales the anisotropy in the interface tension.
(b) We study an active ``crawling'' mechanism, similar to a known mechanism creating convergent extension in the \textit{Xenopus laevis} embryo \cite{Shindo2018}.
To this end, within a cell given cell $c$, we include two pulling forces, which act on the two vertices with minimal and maximal $y$ position, respectively (\autoref{fig:model}b). In any of those two vertices $i$, the pulling force has magnitude $T^\mathrm{act}_\mathrm{crawl}$ and acts along the interface pointing away from the cell $c$ (\autoref{fig:model}b).
To ensure local conservation of linear and angular momentum, we include additional forces that act on the cell centers adjacent to vertex $i$ and cell $c$. In terms of Eq.~\eqref{eq:Fact}, this is equivalent to $N_\mathrm{act}=2N$ active length elements $h_{ci}$ (black double arrows in \autoref{fig:model}b) with tensions $T^\mathrm{act}_{ci}=T^\mathrm{act}_\mathrm{crawl}$.

To numerically solve the time-dependent dynamics, we transform the force balance equation, Eq.~\eqref{eq:force balance}, into a sparse linear problem:
\begin{equation}
	\sum_{j,\beta}{\Gamma_{i\alpha j\beta}} \dot{r}_{j \beta} = F_{i\alpha}^\mathrm{eff}.
	\label{eq:EOM}
\end{equation}
Here, the friction matrix $\Gamma_{i\alpha j\beta}$ and the effective force $F_{i\alpha}^\mathrm{eff}$ depend only on the vertex positions. They are given by:
\begin{align}
	\Gamma_{i\alpha j\beta} &= 
	\sum_{m=1}^{N_\mathrm{fr}} \eta_m\frac{\partial g_m}{\partial r_{i\alpha}} \frac{\partial g_m}{\partial r_{j\beta}},\label{eq:Gamma} \\
	F_{i\alpha}^\mathrm{eff} &= 
	-\frac{\partial E}{\partial r_{i\alpha}}
	-\sum_{m=1}^{N_\mathrm{fr}}{\eta_{m}\frac{\partial g_m}{\partial r_{i\alpha}}\frac{\partial g_m}{\partial\gamma}\dot{\gamma}}
	-\sum_{n=1}^{N_\mathrm{act}}{t_{n}^{a}\frac{\partial h_{n}}{\partial r_{i\alpha}}}
	.
\end{align}
In order for the equations of motion, Eq.~\eqref{eq:EOM}, to have a unique solution, the rank of the friction matrix $\Gamma_{i\alpha j\beta}$ needs to equal the number of degrees of freedom in the system. In other words, there need to be at least $N_\mathrm{fr}\geq2N_v$ frictional elements, where $N_v$ is the number of vertices. For a configuration with only 3-fold vertices, where there are twice as many vertices as cells, this implies $N_\mathrm{fr}\geq4N$. This means that it is not sufficient to frictionally constrain only cell-cell interface lengths, which are only $3N$ elements with $\bar\eta^\ell>0$. We thus also frictionally constrained the $N$ cell areas with $\bar\eta^a>0$. Meanwhile it is not absolutely necessary to have a finite perimeter friction $\bar\eta^p$. We checked that the precise values of these friction coefficients do not affect our results much (\autoref{fig:ap-moduli}).

Note, however, that Eq.~\eqref{eq:EOM} is Galilei invariant, which implies that $\Gamma_{i\alpha j\beta}$ is not fully invertible. We thus include in $\Gamma_{i\alpha j\beta}$ as regularization a very small substrate friction $\eta^s$: $\bm{F}^\mathrm{fr}_{i} = \dots + \eta^s\dot{\bm{r}}_i$. We choose $\eta^s$ small enough to leave the physics unaffected.

For any given simulation run, we initialize the system with a Voronoi tiling of $N=200$ uniformly distributed cell centers, and then minimize the energy $E$.
Starting from this initial configuration, we then solve the equations of motion, Eq~\eqref{eq:EOM}, using an explicit Euler integration scheme with an adaptive time step.
Throughout the manuscript, we choose the parameter values $k_A=10$, $p_0=3.0$, $\bar\eta^a=\bar\eta^p=\bar\eta^\ell=1$, and $\eta^s=10^{-3}$. The T1 cutoff is chosen to be $10^{-2}$. The two activity parameters are varied in the ranges $T_\mathrm{ait}^\mathrm{act}=0\dots2$ and $T^\mathrm{act}_\mathrm{crawl}=0\dots2$. For each parameter set, we run $8\dots 10$ simulations with different random initial configurations.

Generalizing the Batchelor stress formula \cite{Batchelor1970}, we compute the mean-field shear stress $\tilde\sigma_{xx}$ as (see appendix~\ref{app:derivation shear stress formula}):
\begin{equation}
	\tilde\sigma_{xx} = \frac{1}{2N}\Bigg(
	\frac{\partial E}{\partial\gamma} 
	+  \sum_{m=1}^{N_\mathrm{fr}}{\eta_m \dot{g}_m}\frac{\partial g_m}{\partial\gamma}
	+ \sum_{n=1}^{N_\mathrm{act}}{ t_{n}^{a}\frac{\partial h_{n}}{\partial\gamma} }
	\Bigg).
	\label{eq:shear stress}
\end{equation}
Here, the tilde in $\tilde\sigma$ is meant to indicate that we refer to the shear stress, i.e.\ the symmetric, traceless part of the stress tensor.
In the following, we will focus on its $xx$ component exclusively. We will use the simplified notation $\tilde{\sigma}\equiv \tilde{\sigma}_{xx}$.

\subsection{Strain decomposition}
We use the framework from Ref.~\cite{Merkel2017} to decompose the overall tissue shear rate tensor $\tilde{v}_{\alpha\beta}$, i.e.\ the anisotropic part of the strain rate tensor, into contributions by cell shape changes and cell rearrangements:
\begin{equation}
	\tilde{v}_{\alpha\beta} = \frac{\mathrm{D}Q_{\alpha\beta}}{\mathrm{D}t} + R_{\alpha\beta}.
	\label{eq:shear-decomposition}
\end{equation}
Here, $Q_{\alpha\beta}$ is a symmetric, traceless tensor that describes the average cell shape anisotropy, $\mathrm{D}Q_{\alpha\beta}/\mathrm{D}t$ is a co-rotational derivative of that tensor, and $R_{\alpha\beta}$ describes the shear rate contribution by T1 transitions and correlation effects. These correlation effects arise from interactions between local rotation and area expansion with fluctuations in $Q_{\alpha\beta}$ \cite{Merkel2017}.

Importantly, all quantities in Eq.~\eqref{eq:shear-decomposition} can be quantified in the simulations.
Briefly, the overall shear rate $\tilde{v}_{\alpha\beta}$ corresponds to the global tissue deformation imposed through the boundaries of the periodic box \cite{Merkel2014b}.
In this article, we have $\tilde{v}_{xx}=\dot{\gamma}$ and $\tilde{v}_{xy}=0$.
Following Ref.~\cite{Merkel2017}, to quantify average cell shape $Q_{\alpha\beta}$, we first construct a triangulation of the cellular network that corresponds to its dual, where the triangle corners are the cell centers. The shape of each triangle is then characterized by a symmetric, traceless tensor, from which $Q_{\alpha\beta}$ is computed as their area-weighted average \footnote{Different definitions of cell shape are possible, e.g.\ definitions that explicitly depend on the polygonal shape of cells. Yet they usually relate to the triangle-based $Q_{\alpha\beta}$ through a unique non-linear relation, at least approximately. In other words, they contain the same information as the triangle-based $Q_{\alpha\beta}$.}.
Finally, while $R_{\alpha\beta}$ can be computed explicitly from the triangulation \cite{Merkel2017}, here, we compute it using Eq.~\eqref{eq:shear-decomposition}.

In this paper, we focus on pure shear deformation, and thus the $xx$ components of all symmetric, traceless tensors. To simplify notation, we set: $\tilde{v}\equiv \tilde{v}_{xx}\equiv \dot{\gamma}$, $Q\equiv Q_{xx}$, and $R\equiv R_{xx}$.

\section{Results}
In order to obtain a mean-field rheological model for our vertex model, we will focus on the symmetric-traceless part of the stress tensor, relating the anisotropic strain rate, i.e.\ the pure shear rate $\tilde{v}$, to the shear stress $\tilde\sigma$.
Yet, the instantaneous shear stress $\tilde\sigma$ in the vertex model does not only depend on the shear rate, but also on cell shape \cite{Merkel2014b,Lin2023}.
Thus, we aim to construct a shear stress function $\tilde\sigma(Q, \tilde{v})$.

For a complete description of the relation between $\tilde{v}$ and $\tilde\sigma$, the relation $\tilde\sigma(Q, \tilde{v})$ needs to be complemented by an additional relation that allows to predict $Q$.
To this end, we use the shear rate decomposition, Eq.~\eqref{eq:shear-decomposition}, in the form:
\begin{equation}
	\dot{Q} = \tilde{v} - R(Q, \tilde{v}).
	\label{eq:Q dynamics}
\end{equation}
Note that we did not include the corotational term from Eq.~\eqref{eq:shear-decomposition} here, since in our pure shear simulations, there is no average vorticity.
Eq.~\eqref{eq:Q dynamics} can be used to predict the time-dependent $Q$ given the imposed shear rate $\tilde{v}\equiv\dot{\gamma}$, once the rearrangement contribution $R$ is known.
We postulate that $R$ mostly depends on the average $Q$ and the strain rate $\tilde{v}$.
A similar assumption allowed to reconstruct droplet shapes in emulsions relatively well \cite{Guigue2025}.

Because cell shape changes are generally time reversible, while cell rearrangements are not, Eq.~\eqref{eq:Q dynamics} decomposes tissue shear into what we call a \emph{reversible fraction} $\dot{Q}/\tilde{v}$ and an \emph{irreversible fraction} $R/\tilde{v}$, such that $\dot{Q}/\tilde{v} + R/\tilde{v} = 1$ \cite{Guigue2025}.  

In the following, we will first construct the relation $\tilde\sigma(Q, \tilde{v})$ in \autoref{sec:sigma}.
Afterwards, we fit phenomenological relations to $R(Q, \tilde{v})$, which describes yielding in the vertex model, in \autoref{sec:R}.
This approach allows to study the effect of vertex model details on  the shear stress and the yielding behavior separately. Also, it allows us to include cell shape in the rheological equations, which can often be quantified both in simulations and experiments.
Finally, in \autoref{sec:predictions}, we apply our mean-field model to predict cell shape and shear stress in response to large-amplitude oscillatory shear.

\subsection{Constitutive relation for shear stress, $\tilde\sigma(Q, \tilde{v})$}
\label{sec:sigma}

\begin{figure}[t]
	\centering
	\includegraphics[width=\linewidth]{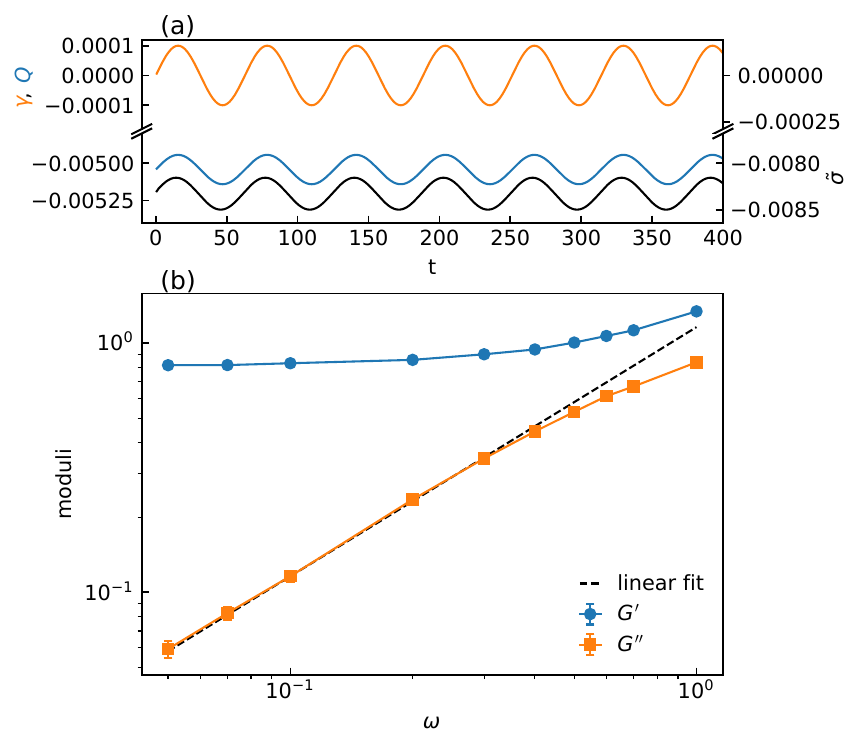}
	\caption{Linear shear rheology.
		(a) Shear protocol, $\gamma(t)$, and corresponding cell shape $Q(t)$ and shear stress $\sigma(t)$. Parameter values: $\gamma_0=10^{-4}$ and $\omega=10^{-1}$. 
		(b) Storage and loss moduli, $G'$ and $G''$, as a function of $\omega$ for strain amplitude $\gamma_0=10^{-7}$.
		Error bars, which are mostly smaller than symbol sizes, show the standard error of the mean.
	}
	\label{fig:osc-shear}
\end{figure}
\subsubsection{Linear visco-elastic regime}
\label{sec:sigma - linear}
We first study the linear regime of $\tilde\sigma(Q, \tilde{v})$, i.e.\ the regime for small $Q$ and $\tilde{v}$, by imposing oscillatory shear, $\gamma(t)=\gamma_{0}\sin(\omega t)$ (\autoref{fig:osc-shear}a, blue curve).
We choose a sufficiently small amplitude so that there are no rearrangements, $R=0$.
Thus, according to Eq.~\eqref{eq:Q dynamics}, $Q(t)=Q_0 + \gamma(t) = Q_0 + \gamma_{0}\sin(\omega t)$, where $Q_0$ is a small average cell shape anisotropy that could be present in the initial, energy-minimized, configuration (shift in $Q$ curve in \autoref{fig:osc-shear}a). 
The shear rate is given by $\tilde{v}\equiv\dot{\gamma}=\gamma_{0}\omega \cos(\omega t)$.
Finally, for small amplitude strain, the shear stress is also sinusoidal with some unknown phase shift, i.e.\ it is the superposition of a sine and a cosine.
Hence, for oscillatory shear with sufficiently small amplitude, the functional form of the time-dependent part of $\tilde\sigma(Q, \tilde{v})$ is known a priori; it has to be of the form:
\begin{equation}
	\tilde\sigma(Q, \tilde{v}) = 2G_0Q + 2\eta_0\tilde{v},
\end{equation}	
where the unknown prefactors $G_0$ and $\eta_0$ are linear shear modulus and shear viscosity, respectively.
The results for $\gamma_{0}=10^{-7}$ are shown in \autoref{fig:osc-shear}b: 
The storage modulus corresponds to the shear modulus, $G'=G_0$, and the loss modulus is proportional to the shear viscosity, $G''=\eta_0\omega$.
We find that both shear modulus and viscosity are constant for small $\omega\lesssim 1$. For larger $\omega$, the shear modulus starts to increase, while the viscosity decreases.
This is consistent with earlier work \cite{Tong2023}, where it has been shown to be due to the internal degrees of freedom not having sufficient time to fully relax during an oscillation period.
Here, we focus on the slow regime, where shear modulus and viscosity are constant, with values $G_0=0.82$ and $\eta_0=1.16$.

\begin{figure}[t]
	\centering
	\includegraphics[width=\linewidth]{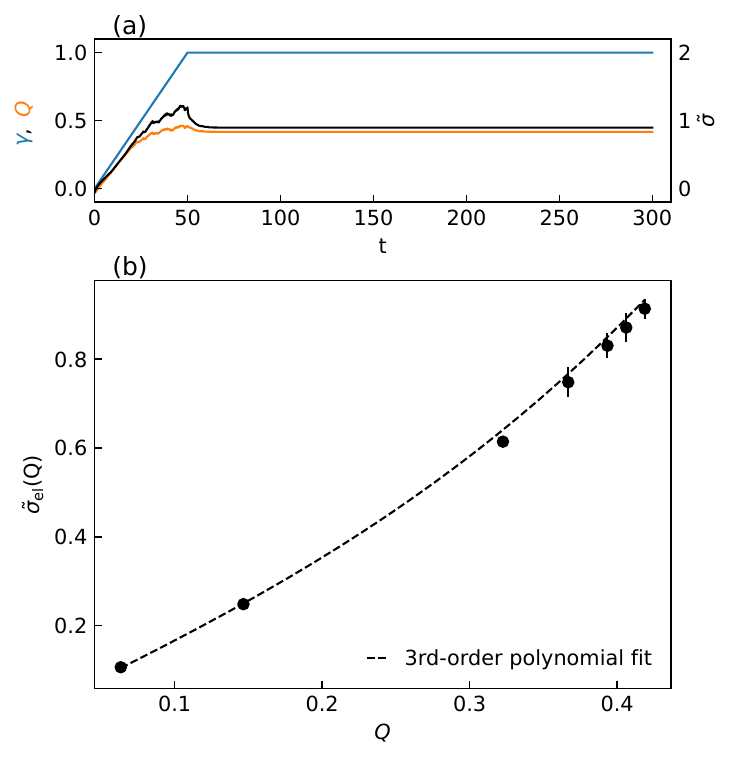}
	\caption{Non-linear elastic regime.
		(a) Shear protocol $\gamma(t)$ and measured $Q$ and $\sigma$.
		(b) Average steady-state stress $\tilde\sigma$ as a function of the average steady-state cell shape $Q$. The dashed line shows a fit to Eq.~\eqref{eq:sigma Q}. Fixing the value $G_0=0.82$ from the linear rheology (\autoref{fig:osc-shear}b), we obtain $G_2=0.86$.
		Error bars show the standard error of the mean.
	}
	\label{fig:stress-Q}
\end{figure}
\subsubsection{Non-linear elastic regime}
\label{sec:sigma - non-linear elastic}
We next extend $\tilde\sigma(Q, \tilde{v})$ to the non-linear elastic regime.
To this end, we carry out simulations with an initial phase with a constant shear rate, $\dot{\gamma}>0$, followed by a relaxation phase with zero shear rate, $\dot{\gamma}=0$ (\autoref{fig:stress-Q}a).
To allow for sufficient pure shear that can be applied to the system, we start from a vertically elongated state with $L_{0x}=10\sqrt{2}/e$ and $L_{0y}=10e\sqrt{2}$, where $e$ is the Euler constant.
We probe the non-linear, elastic behavior of the shear stress, i.e.\ $\tilde\sigma(Q, \tilde{v}=0)$, using 
the final states in these simulations.
Varying the duration of the initial phase and its shear rate allows us to vary $Q$ up to $Q\approx 0.4$.

The results are plotted in \autoref{fig:stress-Q}b.
We find that the data can be fitted to 
\begin{equation}
	\tilde\sigma(Q, \tilde{v}=0)\equiv\tilde\sigma_\mathrm{el}(Q) = 2G_0 Q + 4G_2 Q^3,
	\label{eq:sigma Q}
\end{equation}
where, imposing the known value for $G_0$, we obtain $G_2=0.86$.
In our phenomenological fit function, we only included odd terms, since $\tilde\sigma$ should flip sign whenever $Q$ flips sign.

\subsubsection{Non-linear elasticity with linear viscosity}
\label{sec:sigma - non-linear visco-elastic}
\begin{figure*}[t]
	\centering
	\includegraphics[width=\linewidth]{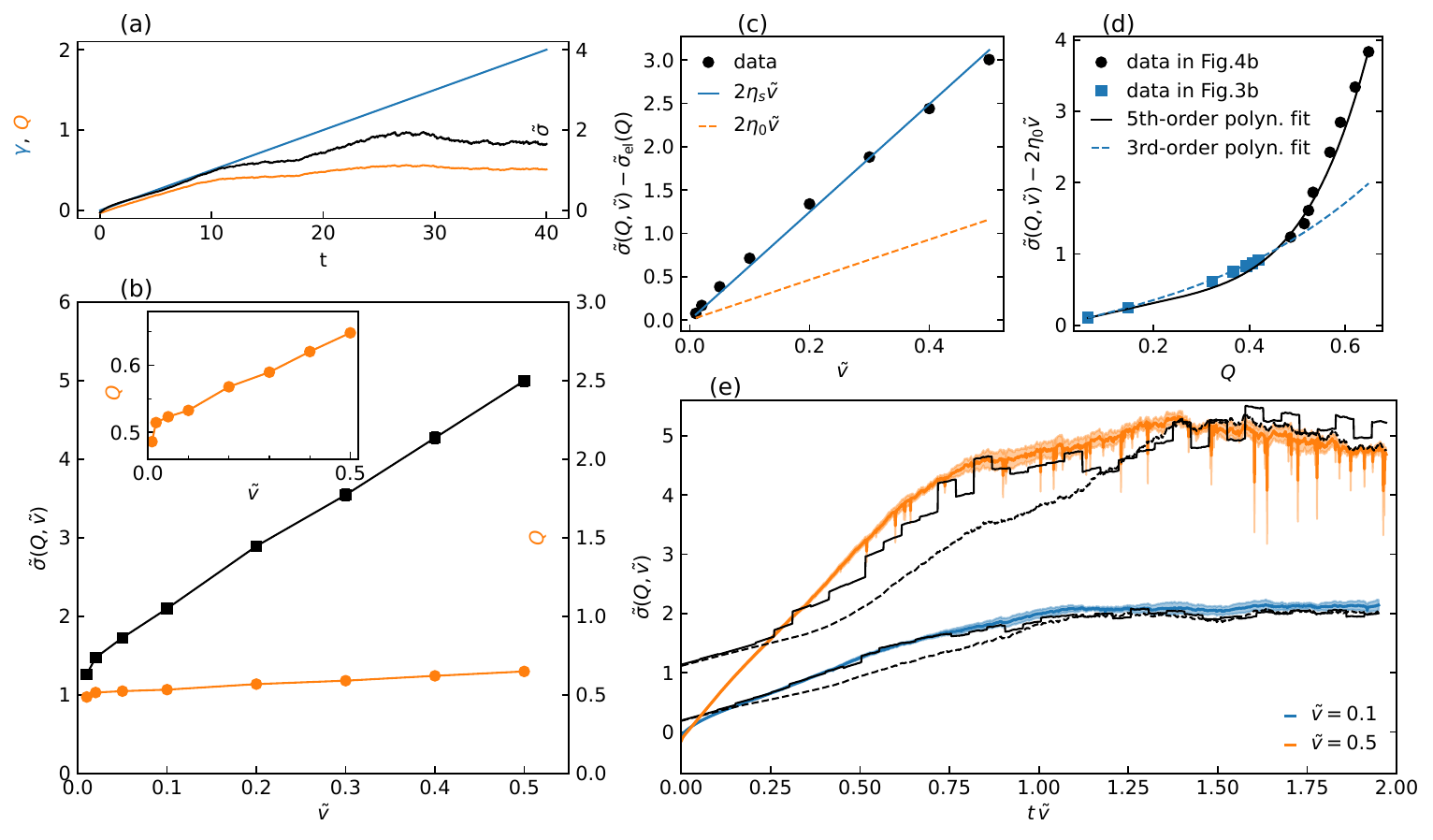}
	\caption{Non-linear elasticity with linear viscosity.
		(a) Shear protocol $\gamma(t)$ and measured $Q$ and $\sigma$.
		(b) Average steady-state stress $\sigma$ and $Q$ as a function of shear rate $\tilde{v}$.
		(b inset) Magnification of steady-state $Q$ over $\tilde{v}$.
		(c) The part of the shear stress $\tilde\sigma$ not captured by the elastic stress $\tilde\sigma_\mathrm{el}$ from \autoref{fig:stress-Q}b increases linearly with the shear rate (black circles). Yet, the associated viscosity $\eta_s$ (blue line) is substantially larger than the viscosity $\eta_0$ obtained from the linear rheology (orange dashed line, cf.\ \autoref{fig:osc-shear}b).
		(d) Conversely, the part of the shear stress $\tilde\sigma$ not captured by the viscous stress expected from the linear rheology, i.e.\ $\tilde\sigma-2\eta_0\tilde{v}$, is larger than the elastic stress function extrapolated from \autoref{fig:stress-Q}b (blue dashed line).
		The blue data points are the same as in \autoref{fig:stress-Q}b and the black data points are those from the constant shear protocol, panel b.
		When taking the union of both data sets and fitting the 5th-order polynomial $\tilde\sigma_\mathrm{el}=2G_0Q + 4G_2Q^3 + 6G_4Q^5$ (black solid line), we obtain $G_2=-0.85$ and $G_4=5.4$ when imposing $G_0=0.82$ from the linear rheology.
		(e) Comparison of the time-dependent shear stress $\tilde\sigma$ measured in the simulations (blue and orange solid lines) with predictions for two different shear rates $\tilde{v}$ (black solid and dashed lines). 
		Solid lines show predictions assuming the scenario in panel c: the elastic stress from Eq.~\eqref{eq:sigma Q} and \autoref{fig:stress-Q}b with an increased viscosity when there are T1 transitions, Eq.~\eqref{eq:viscosity}. To this end, $R(t)$ is computed using a binned average.
		Dashed lines show predictions assuming the scenario from panel d: the elastic stress from the 5th-order polynomial with the linear-rheology viscosity $\eta_0$.
		Shaded regions indicate the standard error of the mean.
	}
	\label{fig:const-protocol}
\end{figure*}

Next, we combine the non-linear elastic regime with a finite shear rate by carrying out simulations at a constant shear rate $\tilde{v}>0$ (\autoref{fig:const-protocol}a).
We again start from the vertically elongated state.

The resulting final values for shear stress and cell shape, averaged over several simulation runs, are plotted in \autoref{fig:const-protocol}b depending on the shear rate $\tilde{v}$.
Probing the dependence of $\tilde\sigma$ on both $Q$ and $\tilde{v}$ is not obvious, since $Q$ and $\tilde{v}$ are correlated in these simulations (\autoref{fig:const-protocol}b inset).
To still make progress, we use a simplified relation for the shear stress, and probe in how far it is consistent with our observations. Specifically, we test whether $\tilde\sigma(Q, \tilde{v})$ can be described by non-linear elasticity acting in parallel with a linear viscosity:
\begin{equation}
	\tilde\sigma(Q, \tilde{v}) \equiv \tilde\sigma_\mathrm{passive}(Q, \tilde{v}) = \tilde\sigma_\mathrm{el}(Q) + 2\eta\tilde{v}.
	\label{eq:sigma Q v}
\end{equation}
Our linear shear rheology points to a rate-dependent shear modulus at large frequencies (\autoref{fig:osc-shear}b). Correspondingly, we expect such a separation to work best for small shear rates.

We first tested whether Eq.~\eqref{eq:sigma Q v} holds using $\tilde\sigma_\mathrm{el}$ from the previous section (Eq.~\eqref{eq:sigma Q}) and  $\eta=\eta_0$ from the linear shear rheology (\autoref{sec:sigma - linear}).
In \autoref{fig:const-protocol}c, we plot $\tilde\sigma(Q, \tilde{v}) - \tilde\sigma_\mathrm{el}(Q)$ over the shear rate $\tilde{v}$, and find indeed a linear relation between both (black filled circles).
However, the slope is much larger than predicted from using $\eta=\eta_0$ (orange dashed line).
In other words, for simulations where the tissue undergoes steady pure shear, the measured shear stress is larger than expected.

There are -- at least -- two possibilities for this deviation from the expected result, either the viscous or the elastic stress was under-estimated.
First, viscous stress may have been under-estimated. Indeed, the blue line in \autoref{fig:const-protocol}c represents a linear fit with much larger viscosity of $\eta_s\approx3.12$.
This would suggest that the viscosity $\eta$ in Eq.~\eqref{eq:sigma Q v} depends on whether there are cell rearrangements in the tissue.

Second, we may have under-estimated the elastic stress, $\tilde\sigma_\mathrm{el}(Q)$.
Indeed, in the previous section, we estimated $\tilde\sigma_\mathrm{el}(Q)$ based on values of $Q\lesssim0.4$ (\autoref{fig:stress-Q}b), which we used to extrapolate the shear stress for $Q\gtrsim0.5$ in this section (\autoref{fig:const-protocol}b).
Thus, maybe higher-order terms in $\tilde\sigma_\mathrm{el}(Q)$ become relevant for $Q\gtrsim0.5$.
To obtain the corresponding $\tilde\sigma_\mathrm{el}(Q)$ in \autoref{fig:const-protocol}d, we plot $\tilde\sigma(Q, \tilde{v}) - 2\eta_0\tilde{v}$ over $Q$, including both the data from the $\tilde{v}=0$ protocol of the previous section (blue squares) and the $\tilde{v}>0$ protocol of this section (black circles).
Taken together, this data can be well described by a fifth-order polynomial for $\tilde\sigma_\mathrm{el}(Q)$ (fit shown as black solid line).

Thus, both hypotheses describe the \emph{stationary-state} data reasonably well.
To distinguish between them, we predict the \emph{time evolution} of the shear stress $\tilde\sigma$ given the known time evolutions of $Q$ and $\tilde{v}$, and compare to the $\tilde\sigma$ measured in the simulations.
To predict $\tilde\sigma$ for the first hypothesis, i.e.\ the shear viscosity is larger when there are T1 transitions, we use Eq.~\eqref{eq:sigma Q v} with $\tilde\sigma_\mathrm{el}(Q)$ from the 3rd-order polynomial fit in \autoref{fig:stress-Q}b in the previous section. Furthermore, to make the viscosity dependent on whether there are T1 transitions, we use the \emph{irreversible fraction} $R/\tilde{v}$ to linearly interpolate between small-amplitude and steady-shear viscosities:
\begin{equation}
	\eta = \eta_0 + \frac{R}{\tilde{v}}(\eta_s - \eta_0).
	\label{eq:viscosity}
\end{equation}
To predict $\tilde\sigma$ for the second hypothesis, i.e.\ the elastic stress becomes much larger beyond $Q\approx 0.4$, we use Eq.~\eqref{eq:sigma Q v} with $\tilde\sigma_\mathrm{el}(Q)$ from the 5th-order polynomial fit in \autoref{fig:const-protocol}d, and we use the viscosity from the linear rheology, $\eta=\eta_0$.
The resulting predictions are shown in \autoref{fig:const-protocol}e by solid (first hypothesis) and dashed (second hypothesis) lines, respectively.
Comparing both predictions to the measured stress (shaded area in \autoref{fig:const-protocol}e), we find that the first hypothesis fits the data better than the second one.
This suggests that the viscosity $\eta$ is strongly increased when the tissue is undergoing T1 transitions.

Note that we find an initial transient deviation between simulation and predictions in \autoref{fig:const-protocol}e:  Both predictions show a finite stress already initially -- this is the expected viscous stress, due to the tissue being sheared at a constant rate.
Yet, in the simulations, the stress is building up over time.
We believe that this stress buildup is related to the small, but finite relaxation time of the internal degrees of freedom \cite{Tong2023}.
This already created the deviation for large frequencies  in the linear shear rheology (\autoref{fig:osc-shear}b), and it is not captured by the simple rheological relations that we aim for in this article.

\begin{figure}%[t]
	\centering
    \includegraphics[width=\linewidth]{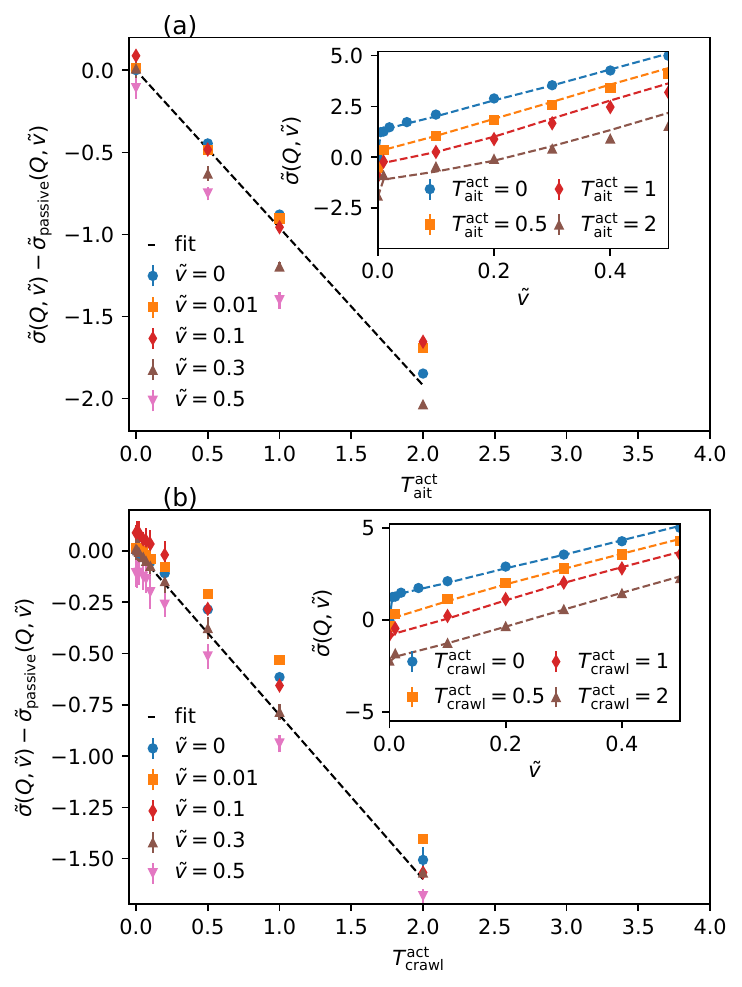}
	\caption{Shear stress created by activity.
		The part of the shear stress $\tilde\sigma$ not captured by the passive stress from Eq.~\eqref{eq:sigma Q v} as a function of the respective  activity parameter for varying shear rate $\tilde{v}$.
		The dashed lines represent linear fits.
		(a) Active, anisotropic interface tensions and
		(b) active crawl forces.
		(insets) The respective total shear stress $\tilde\sigma$ as a function of $\tilde{v}$ for different activity strengths. The dashed lines are the consistency checks using Eqs.~\eqref{eq:sigma with activity ait} and \eqref{eq:sigma with activity crawl}, respectively.
	}
	\label{fig:sig-relation_active}
\end{figure}
\subsubsection{Activity}
We next study how the shear stress $\tilde\sigma(Q, \tilde{v})$ is changed when including two different types of active, anisotropic forces in the vertex model: anisotropic interface tensions, and crawling forces.
To this, end, we carried out constant-shear-rate simulations as in \autoref{sec:sigma - non-linear visco-elastic}, while including either a finite interface tension anisotropy $T_\mathrm{ait}^\mathrm{act}$ or a finite crawl force $T^\mathrm{act}_\mathrm{crawl}$.

We found that in both cases, the activity changed $\tilde\sigma(Q, \tilde{v})$ in a way that shifts the stress curve (see insets to \autoref{fig:sig-relation_active}a,b).
To see how this shift depends on the activity parameter, we plotted $\tilde\sigma(Q, \tilde{v})-\tilde\sigma_\mathrm{passive}(Q, \tilde{v})$ over the activity parameter, respectively, and we found a linear relation in both cases (\autoref{fig:sig-relation_active}a and b).
This suggests the following rheological relations, for active, anisotropic interface tensions:
\begin{equation}
	\tilde\sigma(Q, \tilde{v}) = \tilde\sigma_\mathrm{passive}(Q, \tilde{v}) 
	- \rho_\mathrm{ait}T_\mathrm{ait}^\mathrm{act},
	\label{eq:sigma with activity ait}
\end{equation}
and for an active crawling force
\begin{equation}
	\tilde\sigma(Q, \tilde{v}) = \tilde\sigma_\mathrm{passive}(Q, \tilde{v}) 
	- \rho_\mathrm{crawl}T^\mathrm{act}_\mathrm{crawl},
	\label{eq:sigma with activity crawl}
\end{equation}
where $\tilde\sigma_\mathrm{passive}(Q, \tilde{v})$ is given by Eq.~\eqref{eq:sigma Q v} with elastic stress and viscosity given by Eqs.~\eqref{eq:sigma Q} and \eqref{eq:viscosity}, respectively.
Linear fits yield the values $\rho_\mathrm{ait}=0.96$ and $\rho_\mathrm{crawl}=0.8$ for the activity coefficients (\autoref{fig:sig-relation_active}a and b).

\subsection{Constitutive relation for rearrangements, $R(Q, \tilde{v})$}
\label{sec:R}
\subsubsection{Without activity}
\begin{figure}%[t]
	\centering
	\includegraphics[width=\linewidth]{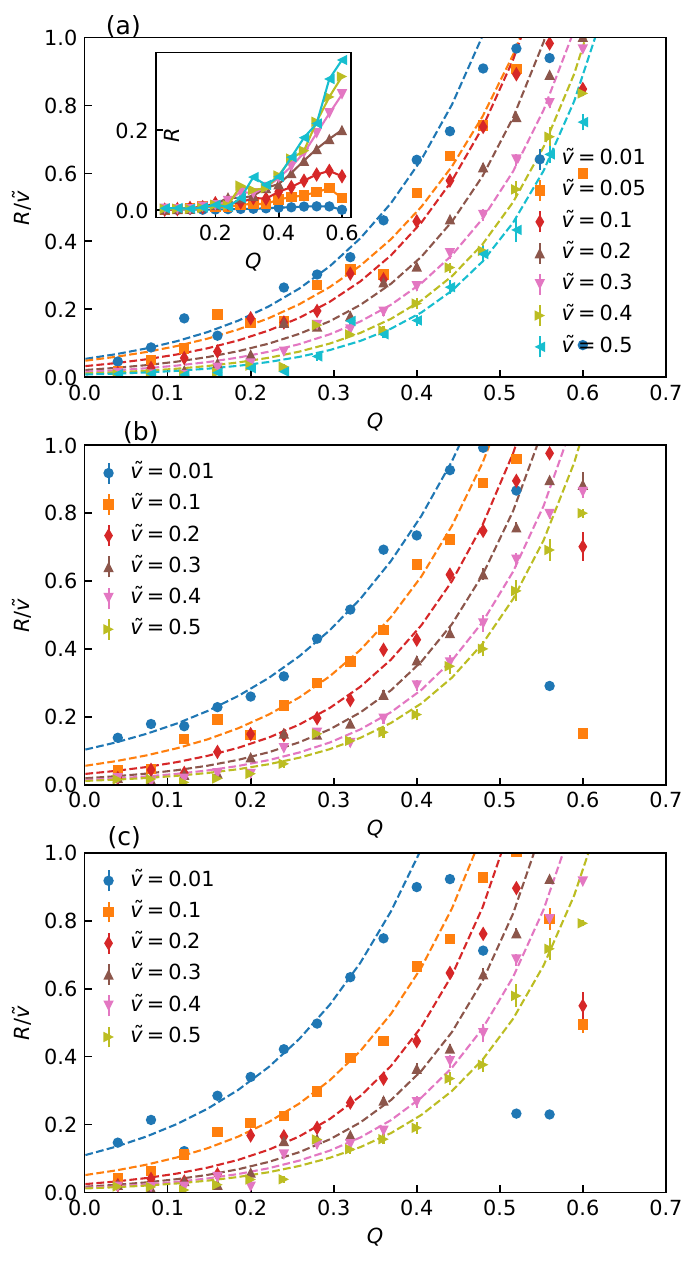}
	\caption{
		Irreversible fraction $R(Q,\tilde{v})/\tilde{v}$ as a function of cell shape $Q$ and shear rate $\tilde{v}$, for 
		(a) passive vertex model tissue,  
		(b) with active, anisotropic interface tension, where $T^{\mathrm{act}}_{\mathrm{ait}}=0.2$, and
		(c) with active crawl force, where $T^{\mathrm{act}}_{\mathrm{crawl}}=0.2$.
	}
	\label{fig:r-relation}
\end{figure}

To probe the dependence of $R$ on $Q$ and $\tilde{v}$, we use the constant-shear-rate simulations from \autoref{sec:sigma - non-linear visco-elastic}.
For each simulation run and time point, we use the $Q$ and $R$ data, and bin them with respect to $Q$, while averaging $R$ within each bin.
The resulting data is shown in \autoref{fig:r-relation}a and inset.

In past work, where sufficiently strong fluctuations were included, $R$ depended mostly only on $Q$ \cite{Merkel2014b,Duclut2021}.
However, this is different in our case, where the function $R(Q)$ strongly depends on the shear rate (\autoref{fig:r-relation}a inset). 
However, the irreversible fraction $R(Q, \tilde{v})/\tilde{v}$ depends much more weakly on $\tilde{v}$ (\autoref{fig:r-relation}a).
This is expected for a regime where the shear rate is smaller than the inherent relaxation times. In other words, we are closer to the quasi-static limit, where $R(Q, \tilde{v})/\tilde{v}$ should become independent of the shear rate \cite{Guigue2025}.

\begin{figure}
	\centering
	\includegraphics[width=\linewidth]{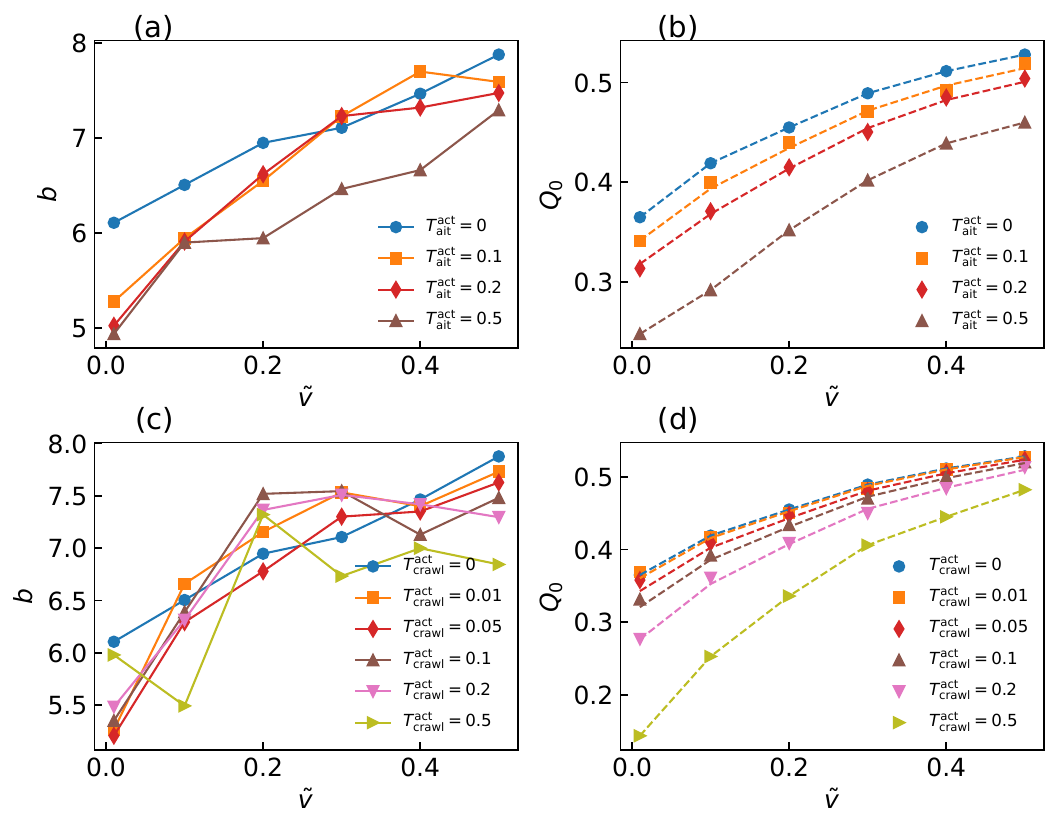}
	\caption{Parameters of the irreversible fraction function fits in \autoref{fig:r-relation} and Eq.~\eqref{eq:irreversible fraction - general}.
	(a,c) $b$ and (b,d) $Q_0$, and their dependence on shear rate $\tilde{v}$ and activity.
	(a,b) For active, anisotropic interface tension. 
	(c,d) For active crawling forces.
	Dashed lines in (b, d) show $Q_0$ computed from Eqs.~\eqref{eq:Q0 - ait} and \eqref{eq:Q0 - crawl} using the $\Delta q_0(\tilde{v})$ functions for activity strength $0.5$, respectively.
	}
	\label{fig:b and Q0}
\end{figure}

We find that for a given shear rate $\tilde{v}$, we can phenomenologically describe the dependence of $R$ on $Q$ by exponentials  (\autoref{fig:r-relation}a, dashed lines):
\begin{equation}
	\frac{R(Q, \tilde{v})}{\tilde{v}} = \frac{1}{2}\exp{\Big(b(\tilde{v})\big[Q-Q_0(\tilde{v})\big]\Big)},
	\label{eq:irreversible fraction - general}
\end{equation}
where the two fit parameters $b$ and $Q_0$ depend on $\tilde{v}$ (\autoref{fig:b and Q0}a and b, blue data points).
Note that, $b$ and $Q_0$ are odd in $\tilde{v}$, i.e.\ $b(-\tilde{v}) = -b(\tilde{v})$ and $Q_0(-\tilde{v}) = -Q_0(\tilde{v})$.
This is because the passive system is invariant with respect to a rotation by $\pi/2$, and thus $R/\tilde{v}$ has to remain the same when simultaneously flipping the signs of $\tilde{v}$ and $Q$. 

\subsubsection{With activity}
When including activity, we find again that $R/\tilde{v}$ can be fitted by exponentials (\autoref{fig:r-relation}b, c), at least as long as the activity is not too large.
Here, we thus focus on small activity parameters, $T_\mathrm{ait}^\mathrm{act}<1$ and $T^\mathrm{act}_\mathrm{crawl}<1$.
We find that the values of $b$ are largely independent of activity (\autoref{fig:b and Q0}a and c), while $Q_0$ \emph{does} depend on activity (\autoref{fig:b and Q0}b and d).
To further simplify the dependence of $Q_0$ on $\tilde{v}$ and activity, we Taylor-expand it in terms of the activity parameters. For active, anisotropic interface tensions:
\begin{equation}
		Q_0(\tilde{v}, T_\mathrm{ait}^\mathrm{act})
		= Q_0^\mathrm{passive}(\tilde{v})
		+ T_\mathrm{ait}^\mathrm{act}\Delta q_0^\mathrm{ait}(\tilde{v}).
		\label{eq:Q0 - ait}
\end{equation}
Note that while $b(\tilde{v})$ and $Q_0^\mathrm{passive}(\tilde{v})$ are odd functions, $\Delta q_0^\mathrm{ait}(\tilde{v})$ has to be an even function.
This is because a rotation by $\pi/2$, which should leave $R/\tilde{v}$ invariant, corresponds to a simultaneous sign flip of $\tilde{v}$, $Q$, and $T_\mathrm{ait}^\mathrm{act}$ (cf.\ Eq.~\eqref{eq:tensions ait}).
Similarly, the relation for the crawl forces read:
\begin{equation}
	Q_0(\tilde{v}, T^\mathrm{act}_\mathrm{crawl})
	= Q_0^\mathrm{passive}(\tilde{v})
	+ T^\mathrm{act}_\mathrm{crawl}\Delta q_0^\mathrm{crawl}(\tilde{v}).
	\label{eq:Q0 - crawl}
\end{equation}
While the symmetry argument cannot be easily made for the active crawling force, in the following we will test the assumption that $\Delta q_0^\mathrm{crawl}(\tilde{v})$ is even as well.
We find that Eqs.~\eqref{eq:Q0 - ait} and \eqref{eq:Q0 - crawl} characterize $Q_0$ quite well, as demonstrated by the dashed lines in \autoref{fig:b and Q0}b and d. These lines were computed using Eqs.~\eqref{eq:Q0 - ait} and \eqref{eq:Q0 - crawl} with $\Delta q_0(\tilde{v})$ taken from the $Q_0$ data at activity strengths $0.5$, respectively.

\begin{figure}%[t]
	\centering
	\includegraphics[width=\linewidth]{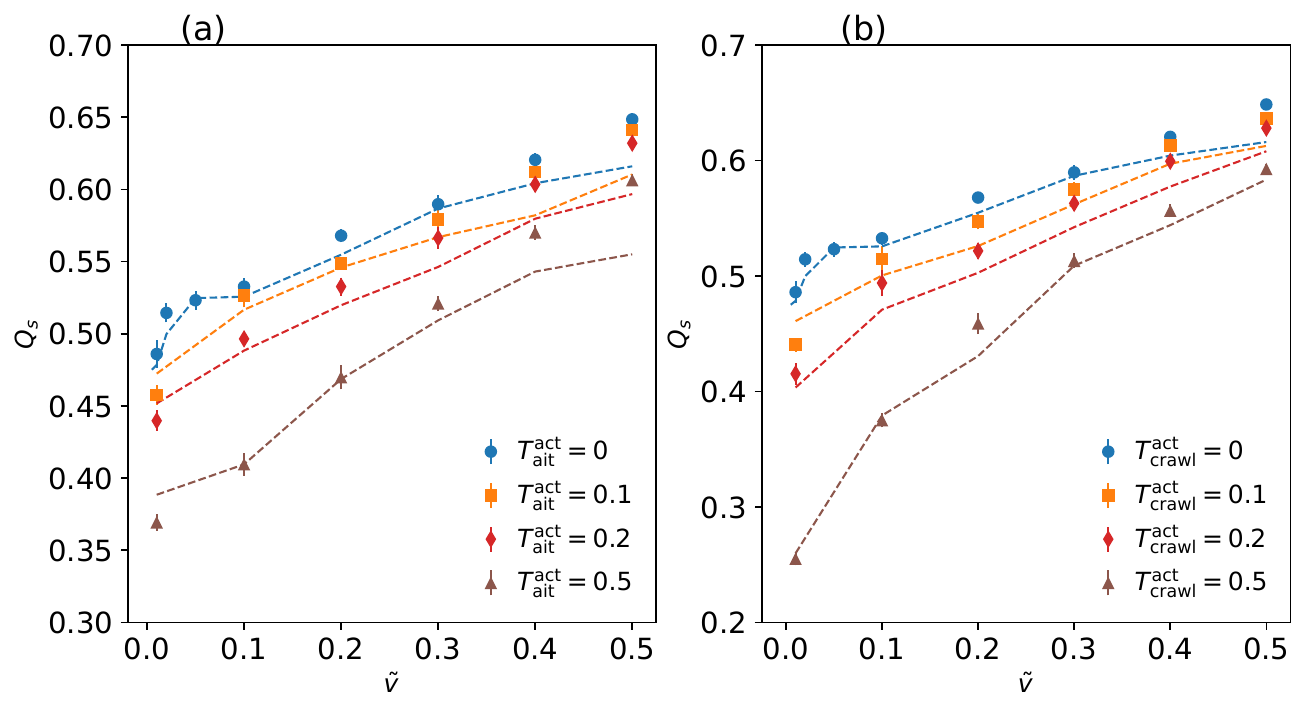}
	\caption{Measured steady-state cell shape $Q_s$ and its prediction as a function of shear rate $\tilde{v}$ for different values of 
		(a) active, anisotropic line tension and 
		(b) active crawling force.
		Closed symbols show the simulation data and dashed lines are the predictions based on Eq.~\eqref{eq:Qs}.
	}
	\label{fig:q-prediction}
\end{figure}

\begin{figure*}[t]
	\centering
	\includegraphics[width=\linewidth]{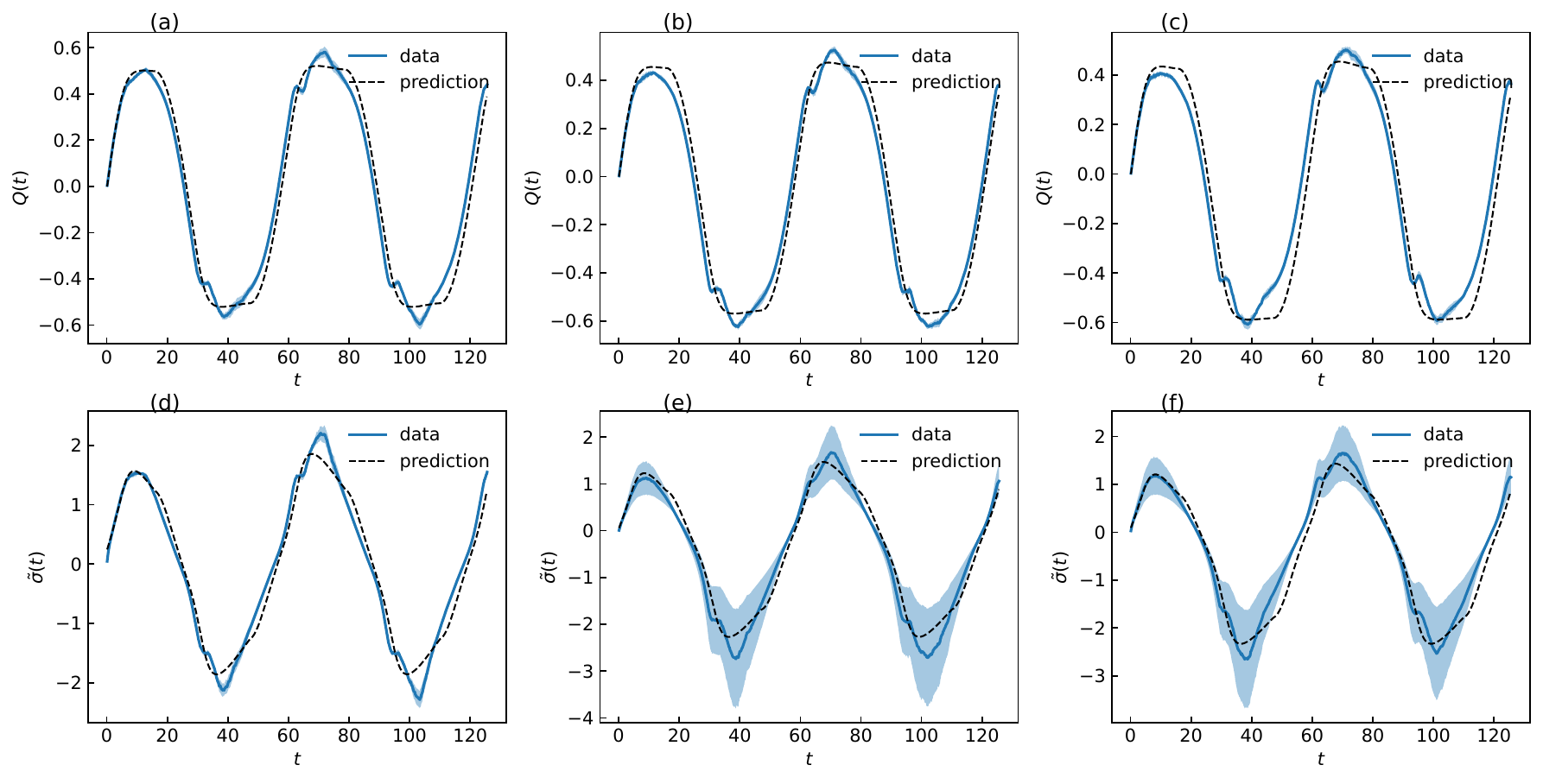}	
	\caption{Predictions for (a,b,c) $Q$ and (d,e,f) $\tilde\sigma$ for large-amplitude oscillatory shear. 
		Blue solid lines show simulation data, while dashed lines show predictions. Shaded regions represent the standard error of the mean.
		(a,d) Passive tissue,
		(b,e) active, anisotropic interface tension at $T_\mathrm{ait}^\mathrm{act}=0.2$, and
		(c,f) active crawling at $T_\mathrm{crawl}^\mathrm{act}=0.2$.
	}
	\label{fig:oscShear-prediction}
\end{figure*}
\subsection{Predictions for cell shape $Q$ and shear stress $\tilde\sigma$}
\label{sec:predictions}
We now use the relations we obtained for $\tilde\sigma$ and $R$ to make predictions for cell shape $Q$ and shear stress $\tilde\sigma$ for some given shear protocol.
Specifically, for $\tilde\sigma$ we use Eqs.~\eqref{eq:sigma Q}--\eqref{eq:sigma with activity crawl}, while for $R$, we use Eqs.~\eqref{eq:irreversible fraction - general}--\eqref{eq:Q0 - crawl} together with the $b(\tilde{v})$, $Q_0^\mathrm{passive}(\tilde{v})$, and $\Delta q_0(\tilde{v})$ data from Fig.~\ref{fig:b and Q0}, taking into account their respective symmetry properties.

\subsubsection{Steady-state cell shape}
First, as a consistency check for the obtained $R$ functions, we predict the steady-state value for cell shape, $Q_s$, for the constant-shear-rate simulations. At long times, under steady shear, all of the shear will be accommodated by cell rearrangements, i.e.\ the irreversible fraction is $R/\tilde{v}=1$. Insertion into Eq.~\eqref{eq:irreversible fraction - general} and inverting yields:
\begin{equation}
	Q_s = Q_0 + \frac{\log{2}}{b}.
	\label{eq:Qs}
\end{equation}
Indeed, comparison to the values obtained in the simulation show an excellent fit for passive tissues, and tissues with active, anisotropic interface tension and active crawling forces (\autoref{fig:q-prediction}a and b).
The increase of $Q_0$ with shear rate (\autoref{fig:b and Q0}b and d) also explains why we observed a larger steady-state cell shape, $Q_s$, in the constant-shear-rate simulations (\autoref{sec:sigma - non-linear visco-elastic}), where $Q_s\gtrsim0.5$, than in the two-phase shear simulations with final shear rate zero (\autoref{sec:sigma - non-linear elastic}), where $Q_s\lesssim0.4$.
Moreover, this also explains why the two active cellular mechanisms can create different steady-state cell shapes: 
For the crawling mechanism, $\Delta q_0$ has a higher negative amplitude than for the anisotropic interface tensions, in particular for small $\tilde{v}$. This translates into smaller $Q_0$, and thus a smaller steady-state value $Q_s$. As a consequence, for larger activity magnitude as used in our earlier work \cite{Barrett2025}, for the crawling mechanism, cell shapes can even become perpendicular to the direction of shear.

\subsubsection{Large-amplitude oscillatory shear}
After this consistency check, we predict the result of a completely different type of simulation -- large-amplitude oscillatory shear.
For frequency $\omega=0.1$ and the amplitude $\gamma_0=1$, we plot the trajectories for $Q(t)$  and $\tilde\sigma(t)$ in Fig.~\ref{fig:oscShear-prediction} for the purely passive case (panels a and d) as well as the two active cases (panels b, c, e, and f).
We find that our predictions (black dashed lines) matched the respective simulated trajectories (blue solid lines) relatively well in all cases.
There are some deviations before the reversal of the shear direction, when the tissue is yielding. This may be due to the fact that we apply our theory gauged from constant-shear-rate simulations to a simulation with a varying shear rate.
The observed deviations may thus arise from some form of memory of the past deformation history that is not fully captured by the average cell shape $Q$ alone.

\section{Discussion}
We have developed a mean-field rheological model for the vertex model.
While this paper is not the first one studying vertex model rheology \cite{Farhadifar2007,Staple2010,Merkel2014b,Bi2015,Moshe2017,Noll2017,Sussman2018,Merkel2018,Merkel2019,Wang2020,Duclut2021,Duclut2022,Huang2022,Hernandez2022,Tong2022a,Grossman2022,Tong2023,Lee2024,Lee2024a,Kim2024,Kim2024,Damavandi2025,Grossman2025,Lin2026}, we note that there are important differences as compared to past work.

First, we follow a systematic approach based on a decomposition of the tissue shear rate $\tilde{v}$ into a \emph{reversible} contribution by change of cell shape $Q$ and an \emph{irreversible} contribution by cell rearrangements $R$ \cite{Merkel2014b,Merkel2017,Guigue2025}.
This allows us to split the question of obtaining a mean-field rheology into two separate relations, one about the rearrangements, $R(Q, \tilde{v})$, and one about the shear stress, $\tilde\sigma(Q, \tilde{v})$.
Also, this automatically allows to include cell shape $Q$ in the rheological relations, which can often be determined more easily in experiments than shear stress \cite{Etournay2015,Guirao2015,Tlili2020,Gomez-Gonzalez2020}.

Second, we studied here a vertex model whose dissipation dynamics fulfills translational and rotational Galilean invariance, i.e.\ with exclusively tissue-internal friction instead of the very common friction with an external substrate.
This is important, since there are many tissues without external substrate. Moreover, it is known that the behavior of active materials can strongly depend on the type of friction, e.g.\ an external friction can suppress a well-known active-matter instability \cite{Simha2002,Voituriez2005,Marchetti2013,Ibrahimi2023,Ibrahimi2025}.
Only few vertex model implementations include a tissue-internal friction without external bulk forces or torques, and, as far as we know, this is the first time the rheology of a vertex model with internal friction is systematically probed.

Finally, while there are several papers studying the \emph{linear} vertex model rheology, or that focus on the elastic behavior only, here we discussed the full active, non-linear visco-elasto-plastic rheology of the vertex model.

In our mean-field rheological model, the apparent vertex model viscosity on long times results from two effects: First, a larger shear rate leads to larger steady-state cell shape $Q_s$, and thus a larger elastic stress $\tilde\sigma_\mathrm{el}$ (cf.\ Eq.~\eqref{eq:sigma Q}). This is the only mechanism that created an effective viscosity in some past work \cite{Merkel2014b,Duclut2021,Duclut2022}. 
Yet, additionally, here we explicitly include internal friction in the vertex model, and in our mean-field description we account for an explicit viscosity $\eta$ (cf.\ Eq.~\eqref{eq:sigma Q v}). 
A key result of our work is that this viscosity strongly depends on whether the system shears visco-elastically due to cell shape changes, or visco-plastically due to cell rearrangements.
In the latter case, the viscosity is three times as large as in the former.
We suspect that the additional shear stress is created by the relaxation of cell shapes right after the occurrence of T1 transitions.

Our work further illustrates that the vertex model rheology strongly depends on the type of vertex model dynamics used.
For instance, in studies where dynamics was created only by line tension fluctuations, a very different yielding behavior was observed, with linear $R(Q)$ for large fluctuation amplitudes \cite{Merkel2014b,Duclut2021}, corresponding to a vertex model behavior as a Maxwell liquid.
Meanwhile, we observe exponential relations $R(Q)$ that also depend on the shear rate $\tilde{v}$, reflecting visco-plastic yielding of the vertex model in the absence of fluctuations.

There are several ways in which our work can be extended.
For instance, we needed to describe the irreversible fraction $R/\tilde{v}$ by some phenomenological fit function, which is currently a purely empirical interpolation of the observed data.
In the future, it will be interesting to study how these curves emerge, for instance using some mesoscopic model.
Such a mesoscopic model may also help refine predictions for non-constant-shear-rate simulations (\autoref{fig:oscShear-prediction}).
Moreover, we occasionally observed partial crystallization, and future work could suppress this using polydisperse cell parameters in the vertex model energy.
Finally, while the rheological model developed here focuses on the mean-field limit, it would be interesting to include gradient terms in such a description to describe the interactions between neighboring tissue regions.

\begin{acknowledgments}
The project leading to this publication was supported by the grant RobustTissue attributed to M.M.\ by the French National Research Agency (ANR-22-CE30-0039).

Following the \href{https://www.coalition-s.org/rights-retention-strategy/}{Rights Retention Strategy of Plan S}, a CC-BY 4.0 public copyright license has been applied by the authors to the present document and will be applied to all subsequent version, including the Author Accepted Manuscript arising from this submission. This does not apply to the Version of Record, for which this paragraph can be removed.
\end{acknowledgments}	

\appendix
\section{Shear stress formula}
\label{app:derivation shear stress formula}
To derive the formula from the shear stress, Eq.~\eqref{eq:shear stress}, we pursue a virtual work approach.
Considering some virtual changes in the box shear, $\delta\gamma$, as well as in the vertex positions, $\delta\bm{r}_i$, we state that all virtual work exerted on the system by the boundaries, $\delta W_\mathrm{ext}$, or by active work $\delta W_\mathrm{act}$ acts to increase the potential energy of the system, $\delta E$, and is dissipated, $\delta W_\mathrm{diss}$:
\begin{equation}
	\delta W_\mathrm{ext} + \delta W_\mathrm{act} = \delta E + \delta W_\mathrm{diss}.
	\label{eq:virtual work balance}
\end{equation}
We have
\begin{align}
	\delta W_\mathrm{ext} &= F^\mathrm{ext}_\alpha\frac{\partial L_\alpha}{\partial\gamma}\delta\gamma	\label{eq:delta W ext}
	\\
	\delta W_\mathrm{act} &= -\sum_{n=1}^{N_a}{t^a_n\delta h_n} \label{eq:delta W act}\\
	\delta W_\mathrm{diss} &= \sum_{m=1}^{N_f}{t^f_m\delta g_m},
		\label{eq:delta W diss}
\end{align}
where $F^\mathrm{ext}_\alpha$ for $\alpha \in \lbrace x, y\rbrace$ denotes the external force acting to expand the box in $\alpha$ direction, and the sum over equal dimension indices is implied. Furthermore, we have:
\begin{align}
	\delta h_n &= \frac{\partial h_n}{\partial\gamma}\delta\gamma + \sum_{i}{\frac{\partial h_n}{\partial r_{i\alpha}}\delta r_{i\alpha}}, \\
	\delta g_m &= \frac{\partial g_m}{\partial\gamma}\delta\gamma +\sum_{i}{\frac{\partial g_m}{\partial r_{i\alpha}}\delta r_{i\alpha}}\label{eq:delta g} \\	
	\delta E &= \frac{\partial E}{\partial\gamma}\delta\gamma - \sum_{i}{F^\mathrm{el}_{i\alpha}\delta r_{i\alpha}}.
	\label{eq:delta E}
\end{align}
Finally, we have:
\begin{align}
	F^\mathrm{ext}_x &= L_y\sigma_{xx} & 	
	F^\mathrm{ext}_y &= L_x\sigma_{yy} \\
	\frac{\partial L_x}{\partial\gamma} &= L_x  & 
	\frac{\partial L_y}{\partial\gamma} &= -L_y.
\end{align}
Taken together, this implies that
\begin{equation}
	\delta W_\mathrm{ext} = L_xL_y(\sigma_{xx}-\sigma_{yy})\delta\gamma = 2N\tilde\sigma_{xx}\delta\gamma \label{eq:delta W ext 2}
\end{equation}
Inserting Eqs.~\eqref{eq:delta W act}--\eqref{eq:delta E} and \eqref{eq:delta W ext 2} into Eq.~\eqref{eq:virtual work balance} and comparing the coefficients in front of $\delta\gamma$, we obtain Eq.~\eqref{eq:shear stress}.

\begin{figure}
	\centering
	\includegraphics[width=\linewidth]{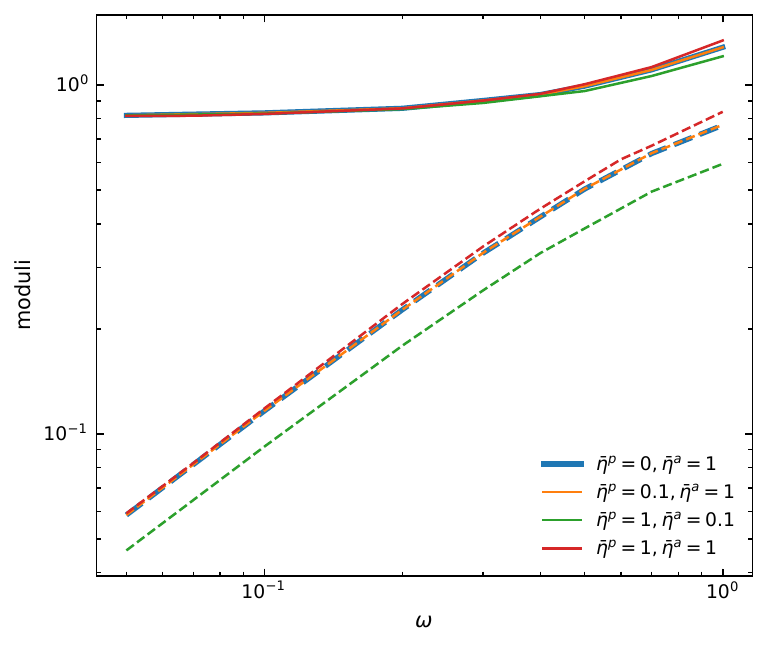}
	\caption{Storage and loss moduli $G'$ and $G''$, as a function of $\omega$ for strain amplitude $\gamma_0=10^{-7}$ for different sets of friction coefficients for the corresponding geometric elements.}
	\label{fig:ap-moduli}
\end{figure}
\section{Role of frictional coefficients}
We studied the effect of varying the individual frictional coefficients $\bar\eta^\ell$, $\bar\eta^a$, and $\bar\eta^p$. Note that one of these coefficients sets the time scale of the system (since all other parameters of our model do not include time), and thus does not need to be varied. Without limiting the generality of the foregoing, we thus fix $\bar\eta^\ell=1$. In \autoref{fig:ap-moduli}, we probe the effect of changing the other two friction parameters, $\bar\eta^a$ and $\bar\eta^p$, on the linear rheology of the system, and we find that it is mostly unchanged, with one exception: Decreasing the area viscosity $\bar\eta^a$ by a factor of 10 slightly decreases the shear viscosity of the vertex model tissue.

\bibliography{references}

\end{document}